\documentclass[twocolumn,superscriptaddress,pra]{revtex4}
\usepackage{doi, graphicx,hyperref, color}
\usepackage{amssymb, amsmath, amsfonts,amsthm,latexsym, dsfont, array, layout}
\usepackage{braket,subfigure, natbib,stackrel}
\usepackage{graphicx,color,xcolor}
\usepackage{booktabs}
\usepackage{array,multirow}

\def\>{\rangle}

\def\<{\langle}

\hypersetup{
    colorlinks=true,  
    linkcolor=blue, 
    citecolor=blue, 
    filecolor=blue,
    urlcolor=blue
}

\begin{document}

\title{Practical quantum somewhat-homomorphic encryption with coherent states}
\author{Si-Hui Tan}
\email{sihui\_tan@sutd.edu.sg}
\affiliation{Singapore University of Technology and Design, 8 Somapah Rd, Singapore 487372}
\affiliation{Centre for 
Quantum Technologies, National University of Singapore, Block S15, 3 Science Drive 2, Singapore 117543, Singapore}
\author{Yingkai Ouyang}
\email{yingkai\_ouyang@sutd.edu.sg}
\affiliation{Singapore University of Technology and Design, 8 Somapah Rd, Singapore 487372}
\affiliation{Centre for 
Quantum Technologies, National University of Singapore, Block S15, 3 Science Drive 2, Singapore 117543, Singapore}
\author{Peter P. Rohde}
\affiliation{Centre for Quantum Software \& Information (QSI), Faculty of Engineering \& Information Technology, University of Technology Sydney, NSW 2007, Australia}

\date{\today}

\begin{abstract}
We present a scheme for implementing homomorphic encryption on coherent states encoded using phase-shift keys. The encryption operations require only rotations in phase space, which commute with computations in the codespace performed via passive linear optics, and with generalized non-linear phase operations that are polynomials of the photon-number operator in the codespace. This encoding scheme can thus be applied to any computation with coherent state inputs, and the computation proceeds via a combination of passive linear optics and generalized non-linear phase operations. An example of such a computation is matrix multiplication, whereby a vector representing coherent state amplitudes is multiplied by a matrix representing a linear optics network, yielding a new vector of coherent state amplitudes. By finding an orthogonal partitioning of the support of our encoded states, we quantify the security of our scheme via the indistinguishability of the encrypted codewords.
Whilst we focus on coherent state encodings, we expect that this phase-key encoding technique could apply to any continuous-variable computation scheme where the phase-shift operator commutes with the computation.
\end{abstract}
\maketitle

\section{Introduction}

In classical cryptography, homomorphic encryption has been a topic of intense interest in recent years \cite{Rivest1978,Gentry:2009:FHE:1536414.1536440,DGH2010}. It is a form of encryption that allows a computation to be performed on the encrypted text without having to first decrypt the text. If an arbitrary computation is allowed, then the encryption is said to be {\it fully homomorphic}. The first fully homomorphic encryption scheme was only discovered recently by Gentry in 2009 \cite{Gentry:2009:FHE:1536414.1536440}. However, like many other classical cryptographic primitives, these homomorphic schemes only offer computational security, which means that they are secure as long as certain problems are computationally intractable. The search for information-theoretically secure encryption problems has led to quantum analogues of homomorphic encryption \cite{PhysRevLett.109.150501, TKOCF, OTF2015}. These schemes only have to hide the quantum input to the computation, unlike a related quantum cryptographic protocol known as blind quantum computation (BQC) \cite{5438603} which also hides the desired computation. However, unlike BQC, no interactive protocols are allowed in quantum homomorphic encryption. Other schemes that perform quantum computing on encrypted data that require interactions are known \cite{Liang2013, Liang2015, FBS2014,Childs:2005:SAQ:2011670.2011674}, though confusingly some of them have been labeled as ``quantum homomorphic encryption" \cite{Liang2013, Liang2015}. Others have focused on hybrid schemes \cite{BJ2015,DSS2016,bib:Alagic2017} that bootstrap on a classical fully homomorphic encryption scheme to achieve computational security while allowing certain classes of quantum computations to be performed on encrypted data. However, some restrictions have arisen. It has been shown that efficient quantum fully-homomorphic encryption is impossible \cite{YPF2014, NS2016}, even when relaxing from perfect to imperfect security. Nonetheless, the key insights contributed by the advent of these quantum schemes still expand the possibilities for implementations of homomorphic encryption in various forms and for different uses, especially since partial information security is still possible for sets of computations of large cardinality \cite{PhysRevLett.109.150501, TKOCF}.

It was shown in \cite{PhysRevLett.109.150501} that homomorphic encryption may be implemented for a restricted class of quantum computation known as the Boson-Sampling model \cite{bib:AaronsonArkhipov10, bib:broome, bib:spring, bib:till, bib:crespi}. In the Boson-Sampling model, computation is performed via a passive linear optical network with a subset of the input modes of this network initialized with a single photon, and the remainder initialized in the vacuum state. To implement the homomorphic encryption described in \cite{PhysRevLett.109.150501}, the client begins by inputting a single-photon into \emph{every} mode, as opposed to just a subset of the modes. Modes where a single photon should have been present are vertically polarized, whereas modes where no photon should have been present are horizontally polarized. Because horizontally and vertically polarized photons do not interfere, they effectively evolve independently through the linear optics network, and by discarding all horizontally polarized photons at the output, the desired computation is recovered. Security is achieved by applying the same random polarization rotation to every photon before entry to the network. The angle of rotation acts as the client's private key, which is not disclosed to the party performing the evaluation. After the evaluation, the photons are returned to the user, who subsequently applies the inverse rotation and discards all horizontally polarized photons, thereby recovering the computation. However, in the absence of knowledge of the key, it is difficult to differentiate between photons that belong to the computation or those which should be discarded. With this scheme, $\mathcal{O}(\log_2(m))$ bits can be hidden when $m$ bits are encrypted. Using group theoretical insights, this homomorphic scheme has been expanded upon to enable quantum computation beyond Boson-Sampling while improving the security \cite{TKOCF} to hide a constant fraction of the number of bits sent. This fraction can be made arbitrarily close to unity by increasing the number of internal states of the bosons used to encode information.

Quantum information theory relies on representing information using quantum states. The underlying algebra of the states, and hence of the operations forming the encoding, affect the performance of the encryption scheme. In this paper, we explore the use of a phase rotation encoding for coherent-state qubits. The advantages of using coherent states are plentiful. Coherent states are produced relatively easily; a laser source closely approximates a coherent state. 
In phase-space a coherent state is a `blob', where the distance from the origin is the amplitude of the coherent state and the angle is its phase. 
Schemes exist to encode classical bits onto coherent states \cite{PhysRevA.68.042319}, to create a collection of universal gate sets for computations with these encodings \cite{PhysRevA.65.042305,PhysRevA.84.050301}, and to map general quantum communication protocols involving pure states of multiple qubits into one that employs coherent states \cite{PhysRevA.90.042335}. The ease of producing, manipulating and distributing coherent states have seeded continuous-variable analogues \cite{RevModPhys.77.513}, primarily featuring coherent states, of quantum cryptography schemes such as quantum key distribution \cite{PhysRevLett.88.057902, GVGWBCG2003,bib:Cao2015} and random ciphers for quantum encryption \cite{PhysRevLett.90.227901, PhysRevA.74.052309}.

In this paper, we present a novel somewhat-homomorphic encryption scheme that utilizes a logical encoding onto coherent states, and encrypts with random rotations in phase space. The scheme works as follows: each classical bit is represented on a single coherent state. A random private key is generated, and the same corresponding random phase shift is applied to every coherent state. An evaluation that is made up of elements from an allowed set of operations, $G$, is then performed on the encrypted data. The set $G$ contains beamsplitters, linear and non-linear phase-shifts, and unitaries that commute with encryption operators. Both the Kerr and cross-Kerr interactions are also included in $G$. In fact, any operator that preserves photon number will work.


We quantify the security of our protocol with the trace distance between any two encrypted inputs. In this notion of security, an adversary without knowledge of the secret key attempts to distinguish the encryptions of any two messages. The smaller the trace distance, the more indistinguishable the encrypted messages are to the adversary. We find this trace distance by showing that the encoding operation induces a partition structure in the states of the microcanonical ensemble where most of the off-diagonal terms are zeroed out. The partition structure gives a closed-form equation for the trace distance between two encrypted inputs. By comparing this trace distance to that for the corresponding unencrypted state, we show that our encryption scheme suppresses the distinguishability of the encoded states and thus provides some security against an adversary attempting to identify the encoded message. Our scheme demonstrates that quantum somewhat-homomorphic encryption is possible for qubit encodings using continuous-variable states. Whilst we focus on a coherent-state encoding, a similar phase-key encoding scheme might be applicable to other continuous-variable (CV) computation schemes.
In principle, this encoding could be applied to any CV scheme where the phase-shift operator commutes with the computation, for any choice of basis states that are not rotation-symmetric in phase-space, such as photon-number states.

\section{Logical encoding using coherent states}Consider an encoding of logical qubits using coherent states with $\ket{0_L}=\ket{\alpha}$, and $\ket{1_L}=\ket{-\alpha}$, where $\ket{\alpha}=\sum_{n=0}^\infty e^{-\frac{|\alpha|^2}{2}}\frac{\alpha^n}{\sqrt{n!}}\ket{n}$ with $\alpha\in\mathbb{C}$. An $m$-bit binary string ${\bf x}:=(x_1,x_2,\ldots, x_m)$ is represented by the tensor product state $\ket{\psi_{\bf x}}=\ket{(-1)^{x_1}\alpha}\ket{(-1)^{x_2}\alpha}\ldots \ket{(-1)^{x_m}\alpha}$. These logical qubits are not orthogonal as $|\braket{\alpha|-\alpha}|^2=e^{-4|\alpha|^2}>0$. Consequently, when $m$ bits are encoded using the ensemble $\{p_{\bf x}, \hat{\rho}_{\bf x}\}$, where $p_{\bf x}$ is the prior probability for the string ${\bf x}$ and $\hat{\rho}_{\bf x}=\ket{\psi_{\bf x}}\bra{\psi_{\bf x}}$, the accessible information of the ensemble, $I_{\rm acc}(\{p_{\bf x},\hat{\rho}_{\bf x}\})$, is less than $m$ bits.

A lower bound on the accessible information of the encoding ensemble can be obtained for a uniform prior by the mutual information between ${\bf x}$ and the outcomes given by a pretty-good measurement (PGM) \cite{Wilde}, ${\bf y}_{\rm PGM}$. The assumption that the prior distribution of the codewords is uniform corresponds to the case where the evaluator has no prior information about the source. The PGM is described by the positive-operator valued measure (POVM) $\{\hat{\rho}^{-\frac{1}{2}}\hat{\rho}_{\bf x}\hat{\rho}^{-\frac{1}{2}}, {\bf x}\in \mathbb{Z}_2^m\}$, where $\hat{\rho}=\frac{1}{2^m}\sum_{{\bf x}\in \mathbb{Z}_2^m}\hat{\rho}_{\bf x}=\frac{1}{2^m}(\hat{\rho}_0+\hat{\rho}_1)^{\otimes m}$, $\hat{\rho}_0:=\ket{\alpha}\bra{\alpha}$, and $\hat{\rho}_1:=\ket{-\alpha}\bra{-\alpha}$. Here $\hat{\rho}^{-\frac{1}{2}}$ denotes the pseudoinverse of the matrix square root of the density matrix $\hat{\rho}$.

Every element of the POVM is a tensor product over the $m$ modes, thus the mutual information for the $m$-mode inputs ${\bf x}$ to outputs ${\bf y}_{\rm PGM}$ is
 \begin{align}\label{eq:accinfoeq1}
 I({\bf x};{\bf y}_{\rm PGM})= m I (x;y_{\rm PGM}) \ , \end{align}
 where the $I(x;y_{\rm PGM})$ is the mutual information for a single-mode discrimination by the PGM. Let $p_x(\ell) :=\frac{1}{2}$ and $p_y(j)$ be the prior and posterior probabilities for obtaining $x=\ell$ and $y=j$ respectively. Then, we have
 \begin{align}\label{eq:accinfoeq2}
 I(x;y_{\rm PGM})=&\sum_{j,\ell=0}^1 p_x(\ell)p(j|\ell)
 \log_2\left(\frac{p(j|\ell)}{p_y(j)}\right )\ ,
 \end{align}
where $p(j|\ell):={\rm tr}(\Pi_j \ket{\ell_L} \bra{\ell_L})$, and $\Pi_j=2(\hat{\rho}_0+\hat{\rho}_1)^{-\frac{1}{2}}\hat{\rho}_{x_j}(\hat{\rho}_0+\hat{\rho}_1)^{-\frac{1}{2}}$ is the conditional probability that the $j$th outcome was measured given that $\ket{\ell_L}$ was sent, and 
\begin{align}
\Pi_j:=\left(\frac{\hat{\rho}_0+\hat{\rho}_1}{2}\right)^{-\frac{1}{2}}\ket{(-1)^j\alpha}\bra{(-1)^j\alpha}\left(\frac{\hat{\rho}_0+\hat{\rho}_1}{2}\right)^{-\frac{1}{2}}\ .
\end{align} 

 The mixed state $\frac{1}{2}(\hat{\rho}_0+\hat{\rho}_1)$ has the spectral decomposition $a_+\ket{\psi_+}\bra{\psi_+}+a_- \ket{\psi_-}\bra{\psi_-}$ \cite{Ban97} where the eigenvectors are 
 \begin{align}
 \ket{\psi_\pm}:=\frac{\ket{\alpha}\pm\ket{-\alpha}}{\sqrt{2}\sqrt{1\pm\exp(-2|\alpha|^2)}} \ ,
 \end{align}
with eigenvalues $a_\pm:=\frac{1}{2}(1\pm\exp(-2|\alpha|^2))$ respectively. The conditional probabilities are explicitly
 \begin{align}
 p(j|\ell)=\left \{
 			\begin{array}{cc}
				\frac{1}{2}\left(\sqrt{a_+}+\sqrt{a_-}\right )^2 \ & , \ j=\ell \\
				\frac{1}{2}\left(\sqrt{a_+}-\sqrt{a_-}\right )^2 \ & , \ j\neq \ell \ ,
			\end{array} 
		\right.
 \end{align}
and thus
\begin{align}
I (x;y_{\rm PGM})=&(\sqrt{a_+}+\sqrt{a_-})^2 \log_2(\sqrt{a_+}+\sqrt{a_-})\nonumber\\
&+ (\sqrt{a_+}-\sqrt{a_-})^2 \log_2(\sqrt{a_+}-\sqrt{a_-})\ .
\end{align}
When $|\alpha|\rightarrow 0$, we have $I (x;y_{\rm PGM})=2 |\alpha|^2/\ln(2)+\mathcal{O}(|\alpha|^4)$, while if $|\alpha|\rightarrow \infty$,  $I (x;y_{\rm PGM})\rightarrow 1$. This is expected because $\ket{\alpha}$ and $\ket{-\alpha}$ are barely distinguishable for small $|\alpha|$, but become nearly orthogonal as $|\alpha|$ becomes large.
 
\section{Homomorphic encryption}
Here, we define encoding and decoding operations that encrypt and decrypt the data. We follow the approach of \cite{TKOCF}, wherein the encoding operators are chosen to commute with those of the computation in the codespace.

After the classical string is encoded onto coherent-state qubits, the user chooses a key $k$ uniformly at random from the set $\{0,1,\ldots, d-1\}$, where $d$ is a positive integer. A phase space rotation is then implemented on every mode, each with the same angle. The phase space rotation operator on the $j$th mode is
\begin{align}
\widehat{\Phi}_j(\theta_k)=\exp(-i\theta_k\hat{a}_j^\dag\hat{a}_j) \ ,
\end{align}
where $\theta_k:=2\pi k/d$. Such an operation on a coherent state yields also a coherent state with the same amplitude, but rotated in phase space by $\theta_k$ around the origin. The application of the above operator on every mode gives a net operator that is generated by the total photon-number operator, $\hat{N}:=\sum_{j=1}^m \hat{a}^\dag_j \hat{a}_j$. The encrypted state is then processed before decryption. The processing is performed by an evaluator, who is able to process the encrypted state without knowing the secret key. Finally, the output bit-string ${\bf y}:=(y_1,y_2,\ldots, y_m)$ can be determined by a measurement on the modes after an inverse rotation $\widehat{\Phi}_j(-\theta_k)$. Since the computation operators are conditioned to commute with the encryption (and decryption) operators and the decryption algorithm is constant in the length of the input, our scheme satisfies the Broadbent and Jeffery's condition of {\it correctness} and {\it compactness} \cite{BJ2015}. In the next section, we will show that non-trivial computation operators which commute with $\bigotimes_{j=1}^m\widehat{\Phi}(\theta_k)$ exist. Then, we discuss the complexity of these allowed 
computations in our scheme. They are closely linked to 
Boson-Sampling \cite{bib:AaronsonArkhipov10} and quantum walks 
\cite{bib:ADZ, bib:RohdeSchreiber10} -- equivalent non-universal models of quantum computation.

\section{Allowed computational operations}The evaluation of quantum operations on the ciphertext is implemented via a unitary operator $U=e^{-iHt/\hbar }$ with its evaluation Hamiltonian $H$ implemented by quantum optical components that are necessarily (Hermitian) photon-number-preserving operators. Using Ehrenfest's Theorem, we have the following evolution of the total photon-number operator $\widehat{N}$ under a given Hamiltonian $H$: $d\braket{\widehat{N}}/dt=\frac{1}{i\hbar}\braket{[\widehat{N},H]}$. Since the evaluation operators do not change the photon-number of the input, then $d\braket{\widehat{N}}/dt=0$. This implies that $\braket{[\widehat{N},H]}=0$. A set of photon-number-preserving computations that also commutes with $\hat{N}$ includes operations in passive linear optics (phase shifts and beamsplitters), and operations that are polynomials of the number operators. We call the latter set the generalized non-linear phase operations and their Hamiltonians are of the form
\begin{align}
H_{\rm NL}:=\sum_{{\bf n}\in \mathbb{N}^m}g_{n_1,\ldots, n_m} \prod_{k=1}^m (a_k^\dag a_k)^{n_k} \ ,
\end{align}
where $g_{n_1,\ldots, n_m}$ is a coupling constant. The single-mode Kerr and cross-Kerr interactions are special cases of $H_{\rm NL}$ \cite{GerryKnight}. Let $K$ be a constant that is proportional to a third-order nonlinear susceptibility. The single-mode Kerr interaction is given by $m=1$, $g_1=-\hbar K$, $g_2=\hbar K$, and $g_{n_1}=0$ otherwise, while the cross-Kerr interaction is given by $m=2$, $g_{1,1}=\hbar K$ and $g_{n_1,n_2}=0$ otherwise.

\begin{figure}[!htb]
\includegraphics[width=0.6\columnwidth]{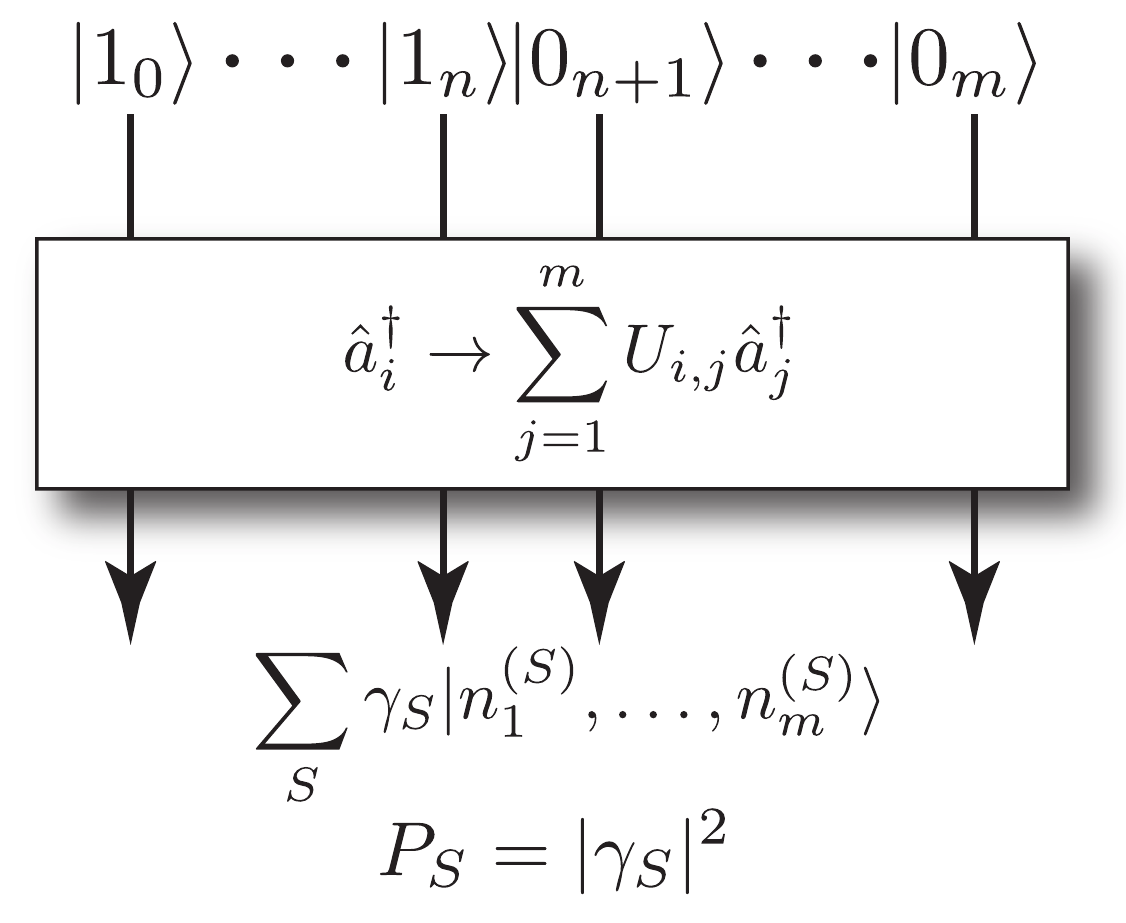}
\caption{The Boson-Sampling model. A string of $n$ single photons is prepared in $m$ optical modes. They are evolved via a passive interferometer $U$. Finally the photon-statistics are sampled from the distribution $P(S)$.} \label{fig:model}
\end{figure}

Passive linear optics is featured heavily in the Boson-Sampling model, where we begin by preparing $n$ single photons in $m$ optical modes (see Fig.~\ref{fig:model}). This input state evolves via non-adaptive, passive linear optics, which implements a unitary map on the photon creation operators, $\hat{a}_i^\dag \to \sum_{j=1}^m U_{i,j} \hat{a}^\dag_j$. The output state to the interferometer has the form, $\ket{\psi_\mathrm{out}} = \sum_S \gamma_S \ket{n_1^{(S)},\dots,n_m^{(S)}}$, where $S$ represents a photon number configuration with $n_i^{(S)}$ photons in the $i$th mode, and $\gamma_S$ are the associated amplitudes. Finally, coincidence photodetection is performed, which samples from the probability distribution $P(S)=|\gamma_S|^2$. Aaronson and Arkhipov showed that sampling from $P(S)$ is likely to be a hard problem for classical computers for some scaling of $m$ with $n$ \cite{bib:AaronsonArkhipov10}. Nonetheless, when the inputs to the circuit are switched from single photons to coherent states, the quantum computation performed can be efficiently simulated classically \cite{PhysRevLett.88.097904}, using simple $m\times m$ matrix multiplication. This changes, however, when we also allow Kerr interactions in the circuit, because this interaction allows the production of cat states from coherent states \cite{YS1986}. For instance, in the interaction picture where $K (\hat{a}^\dag \hat{a})^2$ is regarded as the interaction part of the evaluation Hamiltonian, an initial coherent state will evolve to $e^{-i \hbar K t (\hat{a}^\dag\hat{a})^2}\ket{\alpha}=\frac{1}{\sqrt{2}}\left(e^{-i\pi/4}\ket{\alpha}+e^{i\pi/4}\ket{-\alpha}\right)$ at time $t=\frac{\pi}{2\hbar K}$. Cat states when evolved via passive linear optics and sampled with number-resolved photodetection implements a classically hard sampling problem under plausible complexity theoretic assumptions \cite{RMKFMD15}, although it is not believed to be universal for quantum computation. 


\section{Security Analysis}
Without knowledge of the key, the encrypted input state is
\begin{align}
\mathcal{E}(\hat{\rho}_{\bf x}) &:= 
\frac{1}{d}\sum_{k=0}^{d-1}
	\bigotimes_{j=1}^m\widehat{\Phi}_{j}\left(\frac{2\pi k}{d}\right)
	\ket{\psi_{x_j}}\bra{\psi_{x_j}}
	\widehat{\Phi}_{j}\left(-\frac{2 \pi k}{d}\right)\nonumber\\
	&=\left(\bigotimes_{j=1}^m V_{j}^{x_j} \right)\mathcal{E}(\hat{\rho}_{\bf 0})\left(\bigotimes_{j=1}^m V_{j}^{\dag x_j}\right)\ .\label{eq:microstate}
\end{align}
where $x_j$ is the $j$-th element of the string ${\bf x}$, $\ket{\psi_{x_j}}:=\ket{(-1)^{x_j}\alpha}$ is the state of the $j$th mode of $\ket{\psi_{\bf x}}$, and $V_j=\widehat{\Phi}_j(\pi)$. If someone without knowledge of the key were to attempt to measure the encrypted input state, $\hat{\rho}_{\bf x}$, they would perceive a state highly mixed in the phase degree of freedom, and have difficulty in differentiating between states that belong to the computation. This indistinguishability gives a security for our scheme which we now make precise.

To quantify the security of our encryption scheme, we obtain an upper bound on the trace distance between the encrypted states given by
 $D(\mathcal{E}(\hat{\rho}_{\bf u}), \mathcal{E}(\hat{\rho}_{\bf v}))$ 
for arbitrary pairs of $m$-bit strings {\bf u} and {\bf v},
where $D(\sigma,\tau) = \frac{1}{2} \| \sigma-\tau \|_{\rm tr}$ 
denotes the trace distance between the density matrices $\sigma$ and $\tau$.
It suffices to obtain an upper bound on 
$D(\mathcal{E}(\hat{\rho}_{\bf x}), \mathcal{E}(\hat{\rho}_{\bf 0}))$ where ${\bf x} = {\bf u} \oplus {\bf v}$, because using the invariance of the trace distance under unitary transformation we can get to the trace distance between any pairs of encrypted states.

We first write the phase-shift operator on the Fock space $\widehat{\Phi}\left(\frac{2\pi}{d}\right):=\sum_{y\in\mathbb{N}}\omega^y\ket{y}\bra{y}$
where $\omega=e^{-2\pi i/d}$ and $\mathbb{N}$ is the set of non-negative integers.
Let $\phi({\bf z})=\sum_i z_i \mod d$. Now for every integer $\ell$, the matrix $(\widehat{\Phi}(\frac{2\pi}{d})^{\otimes m})^\ell$ is equivalent to $\sum_{ {\bf y}\in \mathbb{N}^m}\omega^{\ell\phi({\bf y})}\ket{\bf y}\bra{\bf y}$. Hence, using the Fourier identity 
$\frac{1}{d}\sum_{\ell=0}^{d-1}\omega^{\ell\phi({\bf y}-{\bf z})}=\delta_{\phi({\bf y}-{\bf z}),0}$,
\begin{align}\label{eq:E0}
\mathcal{E}(\hat{\rho}_{\bf 0})=& \sum_{{\bf z},{\bf y}\in\mathbb{N}^m}\delta_{\phi({\bf y}-{\bf z}),0}b_{\bf z}b_{\bf y}^*\ket{\bf z}\bra{\bf y} \ ,
\end{align}
where 
$b_{\bf z}=b_{z_1}b_{z_2}\ldots b_{z_m}$ is a product of complex coefficients, each given by $b_n:=e^{-|\alpha|^2/2}\frac{\alpha^n}{\sqrt{n!}}$. The state $\mathcal{E}(\hat{\rho}_{\bf 0})$ admits a block diagonal decomposition, with each block labeled by $\phi({\bf y}-{\bf z})=j$. 
The support of the $j$th block is $\{\ket{{\bf z}}\in G_j: z\in\mathbb{N}^m\}$, where  
$G_j:=\{{\bf z}\in\mathbb{N}^m:\phi({\bf z})=j\}$ 
is a partition of $\mathbb{N}^m$. Defining $\ket{g_j}:=\sum_{{\bf z}\in G_j}b_{\bf z}\ket{\bf z}$, then
\begin{align}\label{eq:block}
\mathcal{E}(\hat{\rho}_{\bf 0})=\sum_{j=0}^{d-1}q_j\ket{\tilde{g}_j}\bra{\tilde{g}_j} \ ,
\end{align}
where $\ket{\tilde{g}_j}=\ket{g_j}/\sqrt{q_j}$ is a normalized state and 
\begin{align}
q_j& =\braket{g_j|g_j}=\sum_{{\bf z}\in G_j}|b_{\bf z}|^2\nonumber \ .
\end{align}

This partition structure makes it straightforward to compute the trace distance between $\mathcal{E}(\hat{\rho}_{\bf x})$ and $\mathcal{E}(\hat{\rho}_{\bf 0})$.
Using eq.~\ref{eq:E0} in the expression in eq.~\ref{eq:microstate}, we have
\begin{align}
\mathcal{E}(\hat{\rho}_{\bf x})= \sum_{\ell=0}^{d-1}q_\ell \ket{\tilde{h}_\ell}\bra{\tilde{h}_\ell} \label{eq:inner1}\ ,
\end{align}
where $\ket{\tilde{h}_\ell}$ is the normalized state
\begin{align}
\ket{\tilde{h}_\ell}=&\bigotimes_{k=1}^m V_k^{x_k}\ket{\tilde{g}_\ell}
= \frac{1}{\sqrt{q_\ell}}\sum_{{\bf z}\in G_\ell}b_{\bf z}(-1)^{{\bf x}\cdot{\bf z}}\ket{{\bf z}} \ .
\end{align}
The states $\ket{\tilde{g}_k}$ and $\ket{\tilde{h}_\ell}$ satisfies the relationship
\begin{align}
\braket{\tilde{h}_\ell|\tilde{g}_k}=\left\{
\begin{array}{cc}
A_k & {\rm if} \ k=\ell\\
0 & {\rm otherwise}
\end{array}
 \right . \ ,
\end{align}
where $A_k=\frac{1}{q_k}\sum_{{\bf z}\in G_k}|b_{\bf z}|^2 (-1)^{{\bf x}\cdot {\bf z}}$ and is a real constant. Owing to the orthogonality of the blocks in the block decomposition of $\mathcal{E}(\hat{\rho}_{\bf 0})$ and $\mathcal{E}(\hat{\rho}_{\bf x})$, we can express the trace distance between them as a sum across blocks. Let $\widehat{O}_k:=\ket{\tilde{h}_k}\bra{\tilde{h}_k}+ \ket{\tilde{g}_k}\bra{\tilde{g}_k}- A_k\ket{\tilde{h}_k}\bra{\tilde{g}_k}- A_k\ket{\tilde{g}_k}\bra{\tilde{h}_k}$. Then
\begin{align}
D(\mathcal{E}(\hat{\rho}_{\bf u}), \mathcal{E}(\hat{\rho}_{\bf v}))
=
\frac{1}{2}\sum_{k=0}^{d-1}q_k{\rm tr}\left(\sqrt{ \widehat{O}_k}\right )= \sum_{k=0}^{d-1}q_k 
\sqrt{1-A_k^2},
\end{align}
where $1-A_k^2$ is the eigenvalue of $\hat{O}_k$ of multiplicity two (please see Appendix \ref{app:eigen} for derivation).

\begin{figure}[t]\centering
        \includegraphics[width=0.9\columnwidth]{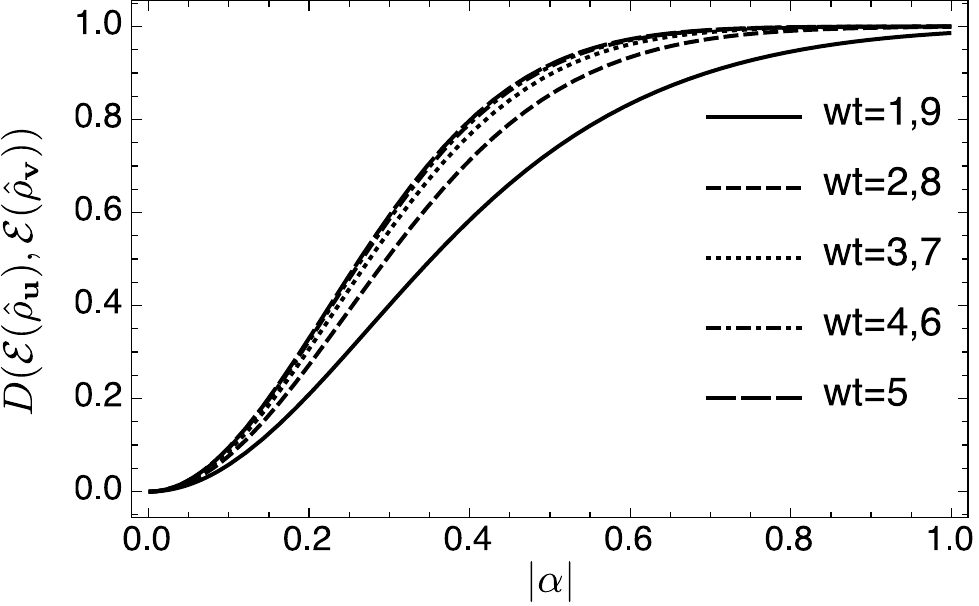}
	\caption{A plot of $D(\mathcal{E}(\hat{\rho}_{\bf u}),\mathcal{E}(\hat{\rho}_{\bf v})) $ versus $|\alpha|$ for $m=10$ and $d=100$, and for various values of $ w ={\rm wt}({\bf u}\oplus {\bf v})$.} \label{fig:tracedist_enc_m10}
	 
\end{figure}

\begin{figure}[t]\centering
        \includegraphics[width=0.9\columnwidth]{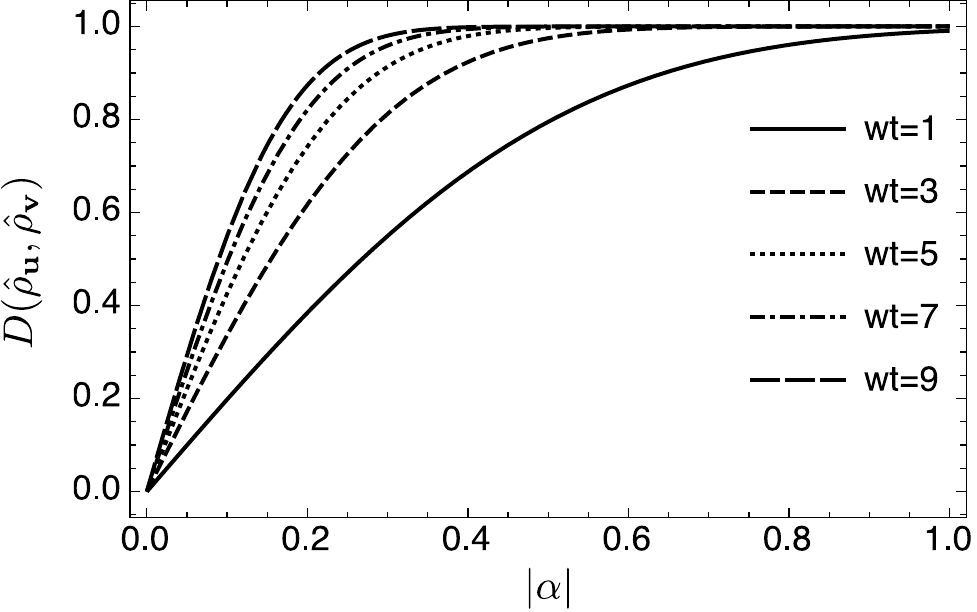}
	\caption{A plot of 
$D(\hat{\rho}_{\bf u},\hat{\rho}_{\bf v})$ versus $|\alpha|$ 
	 for $m=10$ and $d=100$, and for various values of $ w ={\rm wt}({\bf u}\oplus {\bf v})$.} \label{fig:tracedist_unenc_m10}
	 
\end{figure}

In the limit $d\rightarrow \infty$, we can drop the modulus in $\phi({\bf z})$ and use the multinomial theorem to simplify $q_k$ and $A_k$. We have
\begin{align}
q_k\stackrel{d\rightarrow\infty}=e^{-m|\alpha|^2}\frac{(m|\alpha|^2)^k}{k!} \label{eq:qj} \ ,
\end{align}
and
\begin{align}
A_k\stackrel{d\rightarrow\infty}=&\frac{1}{q_k}(m-2{\rm wt}({\bf x}))^k e^{-m |\alpha|^2}\frac{|\alpha|^{2k}}{k!}
\end{align}
respectively, where ${\rm wt}({\bf x})$ is the Hamming weight of ${\bf x}={\bf u}\oplus {\bf v}$. Details of the derivation of $q_k$ and $A_k$ are given in Appendix \ref{app:deriv}. If $d$ is finite, the modulus in the definition of the function $\phi({\bf x})$ prevents us from using the multinomial theorem, and these results would not apply. Explicitly, we have
\begin{align}
D(\mathcal{E}(\hat{\rho}_{\bf v}), \mathcal{E}(\hat{\rho}_{\bf u}))
\stackrel{d\rightarrow\infty}{=}
\sum_{k=1}^{\infty}
\frac{e^{-E}E^k \sqrt{1-\left(\frac{m-2{\rm wt}({\bf x})}{m}\right )^{2k}} }
{k!}
\label{eq:encdist},
\end{align}
where $E=m|\alpha|^2$, and once again ${\bf x}={\bf u}\oplus {\bf v}$.

\begin{figure}[h]\centering
        \includegraphics[width=0.9\columnwidth]{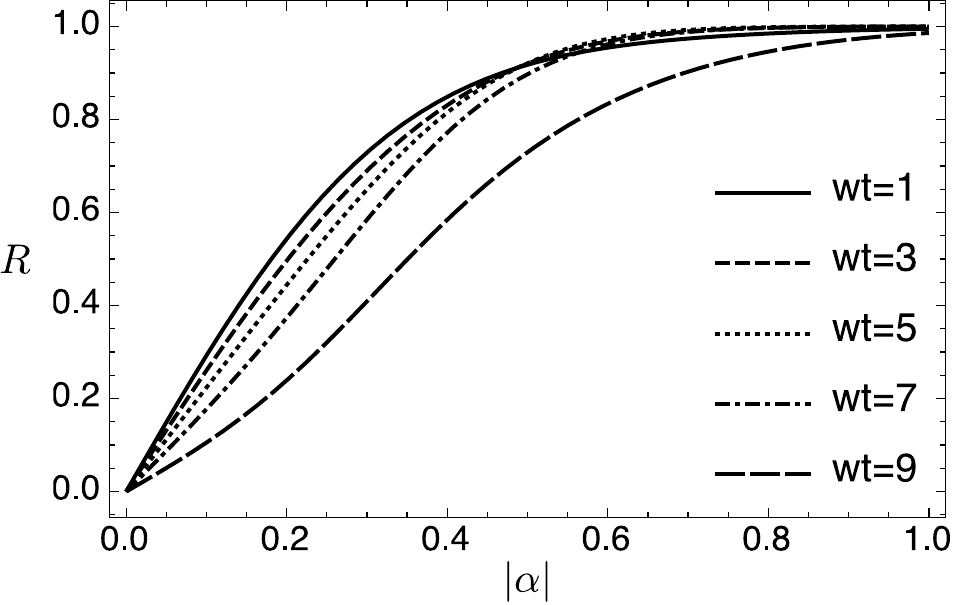}
	\caption{A plot of $R= D(\mathcal{E}(\hat{\rho}_{\bf u}), \mathcal{E}(\hat{\rho}_{\bf v})) / D(\hat{\rho}_{\bf u},\hat{\rho}_{\bf v})$ versus $|\alpha|$ for $m=10$ and $d=100$ for various weights $w = {\rm wt}({\bf u}\oplus {\bf v})$ values. The values of $R$ are less than unity indicating a suppression of distinguishability by the encryption operation.} \label{fig:ratio}
	 
\end{figure}

For comparison, we compute the trace distance between the unencrypted states $\hat{\rho}_{\bf u}$ and $\hat{\rho}_{\bf v}$ which is equal to that between the unencrypted states $\hat{\rho}_{\bf x}$ and $\hat{\rho}_{\bf 0}$ for ${\bf x}={\bf u}\oplus{\bf v}$, because of the invariance of the trace distance under a unitary transformation. This trace distance can be expressed in terms of a rank matrix 
 $\hat{Q}:=\ket{\psi_{\bf x}}\bra{\psi_{\bf x}}+\ket{\psi_{\bf 0}}\bra{\psi_{\bf 0}}-B\ket{\psi_{\bf x}}\bra{\psi_{\bf 0}}-B\ket{\psi_{\bf 0}}\bra{\psi_{\bf x}}$, where $B:=e^{-2{\rm wt}({\bf x})|\alpha|^2}$. Specifically 
\begin{align}\label{eq:unencdist}
D(\hat{\rho}_{\bf u},\hat{\rho}_{\bf v})
=&\frac{1}{2}{\rm tr}\left(\sqrt{\hat{Q}}\right )=\sqrt{1-B^2} \notag\\
=&\sqrt{1-e^{-4 {\rm wt}({\bf x})|\alpha|^2}} \ ,
\end{align}
where $1-B^2$ is the eigenvalue of $\hat{Q}$ (see Appendix \ref{app:eigen} for derivation). The trace distances in eqs.~\ref{eq:encdist} and \ref{eq:unencdist} are plotted for strings of length $m=10$ in Figures \ref{fig:tracedist_enc_m10} and \ref{fig:tracedist_unenc_m10} respectively. Figure \ref{fig:tracedist_enc_m10} was calculated using an encryption key with $d=100$.

The qualitative behaviours of the trace distances with and without encryption are quite similar, with the trace distance vanishing as $|\alpha|\rightarrow 0$, while approaching its maximum value of unity as $|\alpha|$ grows. However, quantitatively, the trace distances are suppressed for the encrypted states (see Figure \ref{fig:ratio}) and have a lower spread over the different ${\rm wt}({\bf x})$ values. The encryption would make it harder for an adversary to distinguish between the different encoded states, thus providing some modest security.

\begin{figure}[h]\centering
        \includegraphics[width=0.9\columnwidth]{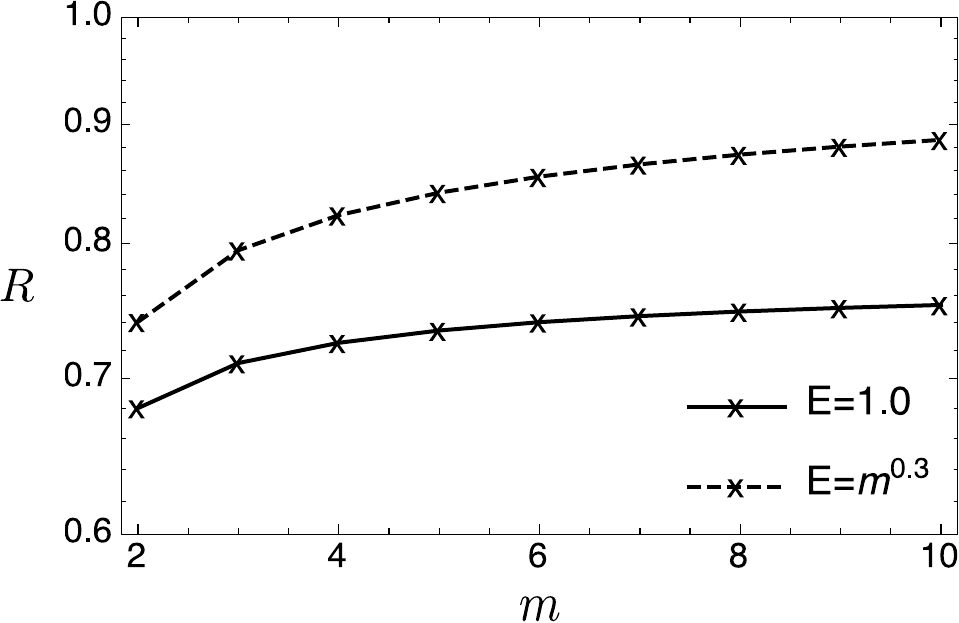}
	\caption{A plot of $R$ versus $m$ with fixed ${\rm wt}({\bf x})=1$ strings, where ${\bf x} = {\bf u} \oplus {\bf v}$, and $d=100$ for (i) $E=1.0$, and (ii) $E=m^r$, where $r=0.3$.} \label{fig:ratios}
\end{figure}

Let $R:= D(\mathcal{E}(\hat{\rho}_{\bf u}), \mathcal{E}(\hat{\rho}_{\bf v})) / D(\hat{\rho}_{\bf u},\hat{\rho}_{\bf v})$, and $E:=m|\alpha|^2$. We plot $R$ versus $m$ for (i) $E=1.0$, and (ii) $E=m^r$, where $r=0.3$ in Fig.~\ref{fig:ratios}.  The ratios are less than unity indicating that the trace distances are suppressed for the encrypted states. However, as the ratios increase with $m$, this suppression diminishes with an increasing length of the encoded string in both energy regimes.

\begin{figure}[h]\centering
        \includegraphics[width=0.9\columnwidth]{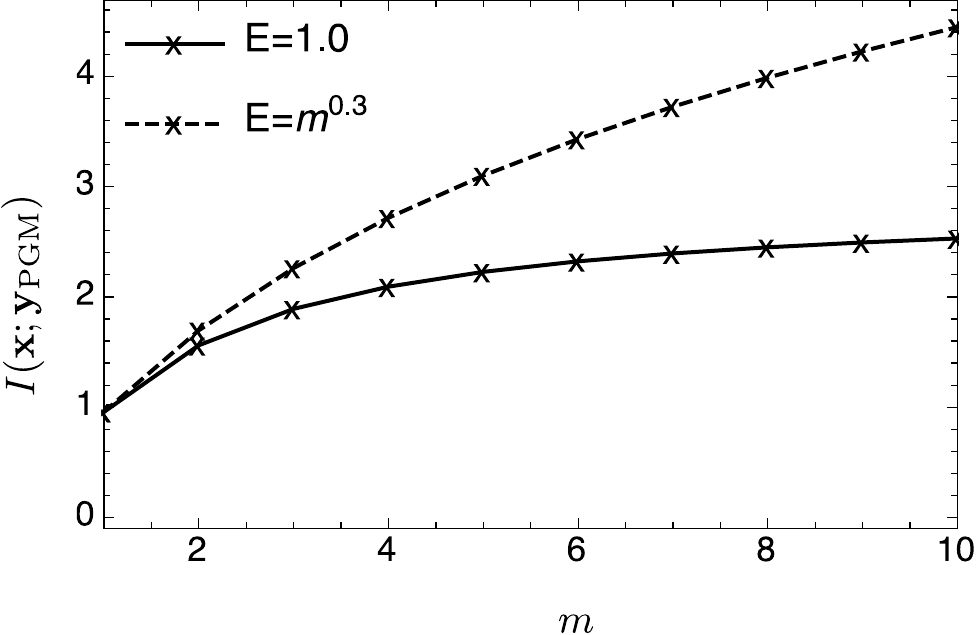}
	\caption{A plot of $I({\bf x};{\bf y}_{\rm PGM})$ versus $m$ with $d=100$, and fixed ${\rm wt}({\bf x})=1$ strings for (i) $E=1.0$ and (ii) $E=m^r$, where $r=0.3$.} \label{fig:accI}
\end{figure}

The corresponding lower bounds on $I({\bf x}; {\bf y}_{\rm PGM})$ are plotted in Fig.~\ref{fig:accI}. It shows $I({\bf x}; {\bf y}_{\rm PGM})$ increasing with $m$ for both (i) $E=1.0$, and $E=m^r$ where $r=0.3$. This means that in these regimes of $E$, someone with the secret key can send more information with increasing code length.

One might hope for an energy regime in which $I({\bf x};{\bf y}_{\rm PGM})$ increases, while the ratio $R$ vanishes with increasing $m$. However, this does not seem to be possible. Our scheme is still useful in situations where secure delegated quantum processing is desired when constrained to preparing simple resources like coherent states, and to short codewords.

\section{Conclusion}
In this paper, we present a new homomorphic encryption scheme that allows processing on logical qubits encoded onto coherent states while encrypted by a random rotation in phase-space. Although the input states are classical, the set of allowed quantum operations is hard to simulate classically. We analyzed the security of our scheme through the trace distance of any two encrypted codewords and showed that there exists regimes of coherent-state amplitudes and bit-string length in which the trace distance can be suppressed indicating increased security afforded by the encryption. Our scheme is readily implementable with existing optical network technology, and is useful as a new primitive for secure delegated quantum computing using continuous-variable resources.

\begin{acknowledgments}
The authors would like to thank J.F.~Fitzsimons for useful discussions. This research was supported in part by the Singapore
National Research Foundation under NRF Award No. NRF-NRFF2013-01.
This research was conducted by the Australian Research Council Centre of Excellence for Engineered Quantum Systems (Project number CE110001013). PPR acknowledges financial support from Lockheed Martin, and is funded by an ARC Future Fellowship (project FT160100397). ST acknowldeges support from the Air Force Office of Scientific Research under AOARD grant FA2386-15-1-4082.
\end{acknowledgments}

\bibliography{paperbib}

\begin{thebibliography}{39}
\expandafter\ifx\csname natexlab\endcsname\relax\def\natexlab#1{#1}\fi
\expandafter\ifx\csname bibnamefont\endcsname\relax
  \def\bibnamefont#1{#1}\fi
\expandafter\ifx\csname bibfnamefont\endcsname\relax
  \def\bibfnamefont#1{#1}\fi
\expandafter\ifx\csname citenamefont\endcsname\relax
  \def\citenamefont#1{#1}\fi
\expandafter\ifx\csname url\endcsname\relax
  \def\url#1{\texttt{#1}}\fi
\expandafter\ifx\csname urlprefix\endcsname\relax\def\urlprefix{URL }\fi
\providecommand{\bibinfo}[2]{#2}
\providecommand{\eprint}[2][]{\url{#2}}

\bibitem[{\citenamefont{Rivest et~al.}(1978)\citenamefont{Rivest, Adleman, and
  Dertouzos}}]{Rivest1978}
\bibinfo{author}{\bibfnamefont{R.~L.} \bibnamefont{Rivest}},
  \bibinfo{author}{\bibfnamefont{L.}~\bibnamefont{Adleman}}, \bibnamefont{and}
  \bibinfo{author}{\bibfnamefont{M.~L.} \bibnamefont{Dertouzos}},
  \bibinfo{journal}{Foundations of Secure Computation, Academia Press} pp.
  \bibinfo{pages}{169--179} (\bibinfo{year}{1978}).

\bibitem[{\citenamefont{Gentry}(2009)}]{Gentry:2009:FHE:1536414.1536440}
\bibinfo{author}{\bibfnamefont{C.}~\bibnamefont{Gentry}}, in
  \emph{\bibinfo{booktitle}{Proceedings of the Forty-first Annual ACM Symposium
  on Theory of Computing}} (\bibinfo{publisher}{ACM}, \bibinfo{address}{New
  York, NY, USA}, \bibinfo{year}{2009}), STOC '09, pp.
  \bibinfo{pages}{169--178}, ISBN \bibinfo{isbn}{978-1-60558-506-2},
  \urlprefix\url{http://doi.acm.org/10.1145/1536414.1536440}.

\bibitem[{\citenamefont{van Dijk et~al.}(2010)\citenamefont{van Dijk, Gentry,
  Halevi, and Vaikuntanathan}}]{DGH2010}
\bibinfo{author}{\bibfnamefont{M.}~\bibnamefont{van Dijk}},
  \bibinfo{author}{\bibfnamefont{C.}~\bibnamefont{Gentry}},
  \bibinfo{author}{\bibfnamefont{S.}~\bibnamefont{Halevi}}, \bibnamefont{and}
  \bibinfo{author}{\bibfnamefont{V.}~\bibnamefont{Vaikuntanathan}}, in
  \emph{\bibinfo{booktitle}{Advances in Cryptology EUROCRYPT 2010}}, edited by
  \bibinfo{editor}{\bibfnamefont{H.}~\bibnamefont{Gilbert}}
  (\bibinfo{publisher}{Springer Berlin Heidelberg}, \bibinfo{year}{2010}), vol.
  \bibinfo{volume}{6110} of \emph{\bibinfo{series}{Lecture Notes in Computer
  Science}}, pp. \bibinfo{pages}{24--43},
  \urlprefix\url{http://dx.doi.org/10.1007/978-3-642-13190-5_2}.

\bibitem[{\citenamefont{Rohde et~al.}(2012)\citenamefont{Rohde, Fitzsimons, and
  Gilchrist}}]{PhysRevLett.109.150501}
\bibinfo{author}{\bibfnamefont{P.~P.} \bibnamefont{Rohde}},
  \bibinfo{author}{\bibfnamefont{J.~F.} \bibnamefont{Fitzsimons}},
  \bibnamefont{and}
  \bibinfo{author}{\bibfnamefont{A.}~\bibnamefont{Gilchrist}},
  \bibinfo{journal}{Phys. Rev. Lett.} \textbf{\bibinfo{volume}{109}},
  \bibinfo{pages}{150501} (\bibinfo{year}{2012}),
  \urlprefix\url{http://link.aps.org/doi/10.1103/PhysRevLett.109.150501}.

\bibitem[{\citenamefont{Tan et~al.}(2016)\citenamefont{Tan, Kettlewell, Ouyang,
  Chen, and Fitzsimons}}]{TKOCF}
\bibinfo{author}{\bibfnamefont{S.-H.} \bibnamefont{Tan}},
  \bibinfo{author}{\bibfnamefont{J.~A.} \bibnamefont{Kettlewell}},
  \bibinfo{author}{\bibfnamefont{Y.}~\bibnamefont{Ouyang}},
  \bibinfo{author}{\bibfnamefont{L.}~\bibnamefont{Chen}}, \bibnamefont{and}
  \bibinfo{author}{\bibfnamefont{J.~F.} \bibnamefont{Fitzsimons}},
  \bibinfo{journal}{Sci.~Rep.} \textbf{\bibinfo{volume}{6}},
  \bibinfo{pages}{33467} (\bibinfo{year}{2016}).

\bibitem[{\citenamefont{Ouyang et~al.}(2015)\citenamefont{Ouyang, Tan, and
  Fitzsimons}}]{OTF2015}
\bibinfo{author}{\bibfnamefont{Y.}~\bibnamefont{Ouyang}},
  \bibinfo{author}{\bibfnamefont{S.-H.} \bibnamefont{Tan}}, \bibnamefont{and}
  \bibinfo{author}{\bibfnamefont{J.}~\bibnamefont{Fitzsimons}},
  \bibinfo{journal}{arXiv:1508.00938}  (\bibinfo{year}{2015}).

\bibitem[{\citenamefont{Broadbent et~al.}(2009)\citenamefont{Broadbent,
  Fitzsimons, and Kashefi}}]{5438603}
\bibinfo{author}{\bibfnamefont{A.}~\bibnamefont{Broadbent}},
  \bibinfo{author}{\bibfnamefont{J.}~\bibnamefont{Fitzsimons}},
  \bibnamefont{and} \bibinfo{author}{\bibfnamefont{E.}~\bibnamefont{Kashefi}},
  in \emph{\bibinfo{booktitle}{Foundations of Computer Science, 2009. FOCS '09.
  50th Annual IEEE Symposium on}} (\bibinfo{year}{2009}), pp.
  \bibinfo{pages}{517--526}, ISSN \bibinfo{issn}{0272-5428}.

\bibitem[{\citenamefont{Liang}(2013)}]{Liang2013}
\bibinfo{author}{\bibfnamefont{M.}~\bibnamefont{Liang}},
  \bibinfo{journal}{Quantum Information Processing}
  \textbf{\bibinfo{volume}{12}}, \bibinfo{pages}{3675} (\bibinfo{year}{2013}),
  ISSN \bibinfo{issn}{1570-0755},
  \urlprefix\url{http://dx.doi.org/10.1007/s11128-013-0626-5}.

\bibitem[{\citenamefont{Liang}(2015)}]{Liang2015}
\bibinfo{author}{\bibfnamefont{M.}~\bibnamefont{Liang}},
  \bibinfo{journal}{Quantum Information Processing}
  \textbf{\bibinfo{volume}{14}}, \bibinfo{pages}{2749} (\bibinfo{year}{2015}),
  ISSN \bibinfo{issn}{1573-1332},
  \urlprefix\url{http://dx.doi.org/10.1007/s11128-015-1034-9}.

\bibitem[{\citenamefont{Fisher et~al.}(2014)\citenamefont{Fisher, Broadbent,
  Shalm, Yan, Lavoie, Prevedel, Jennewein, and Resch}}]{FBS2014}
\bibinfo{author}{\bibfnamefont{K.~A.~G.} \bibnamefont{Fisher}},
  \bibinfo{author}{\bibfnamefont{A.}~\bibnamefont{Broadbent}},
  \bibinfo{author}{\bibfnamefont{L.~K.} \bibnamefont{Shalm}},
  \bibinfo{author}{\bibfnamefont{Z.}~\bibnamefont{Yan}},
  \bibinfo{author}{\bibfnamefont{J.}~\bibnamefont{Lavoie}},
  \bibinfo{author}{\bibfnamefont{R.}~\bibnamefont{Prevedel}},
  \bibinfo{author}{\bibfnamefont{T.}~\bibnamefont{Jennewein}},
  \bibnamefont{and} \bibinfo{author}{\bibfnamefont{K.~J.} \bibnamefont{Resch}},
  \bibinfo{journal}{Nat Commun} \textbf{\bibinfo{volume}{5}}
  (\bibinfo{year}{2014}), \urlprefix\url{http://dx.doi.org/10.1038/ncomms4074}.

\bibitem[{\citenamefont{Childs}(2005)}]{Childs:2005:SAQ:2011670.2011674}
\bibinfo{author}{\bibfnamefont{A.~M.} \bibnamefont{Childs}},
  \bibinfo{journal}{Quantum Info. Comput.} \textbf{\bibinfo{volume}{5}},
  \bibinfo{pages}{456} (\bibinfo{year}{2005}), ISSN \bibinfo{issn}{1533-7146},
  \urlprefix\url{http://dl.acm.org/citation.cfm?id=2011670.2011674}.

\bibitem[{\citenamefont{Broadbent and Jeffery}(2015)}]{BJ2015}
\bibinfo{author}{\bibfnamefont{A.}~\bibnamefont{Broadbent}} \bibnamefont{and}
  \bibinfo{author}{\bibfnamefont{S.}~\bibnamefont{Jeffery}},
  \bibinfo{journal}{Advances in Cryptology CRYPTO 2015}
  \textbf{\bibinfo{volume}{9216}}, \bibinfo{pages}{609} (\bibinfo{year}{2015}),
  \urlprefix\url{http://link.springer.com/chapter/10.1007%2F978-3-662-48000-7_30}.

\bibitem[{\citenamefont{Dulek et~al.}(2016)\citenamefont{Dulek, Schaffner, and
  Speelman}}]{DSS2016}
\bibinfo{author}{\bibfnamefont{Y.}~\bibnamefont{Dulek}},
  \bibinfo{author}{\bibfnamefont{C.}~\bibnamefont{Schaffner}},
  \bibnamefont{and} \bibinfo{author}{\bibfnamefont{F.}~\bibnamefont{Speelman}},
  \bibinfo{journal}{arXiv:1603.09717v1}  (\bibinfo{year}{2016}).

\bibitem[{\citenamefont{Alagic et~al.}(2017)\citenamefont{Alagic, Dulek,
  Schaffner, and Speelman}}]{bib:Alagic2017}
\bibinfo{author}{\bibfnamefont{G.}~\bibnamefont{Alagic}},
  \bibinfo{author}{\bibfnamefont{Y.}~\bibnamefont{Dulek}},
  \bibinfo{author}{\bibfnamefont{C.}~\bibnamefont{Schaffner}},
  \bibnamefont{and} \bibinfo{author}{\bibfnamefont{F.}~\bibnamefont{Speelman}},
  \bibinfo{journal}{arXiv:1708.09156v1}  (\bibinfo{year}{2017}).

\bibitem[{\citenamefont{Yu et~al.}(2014)\citenamefont{Yu, P\'erez-Delgado, and
  Fitzsimons}}]{YPF2014}
\bibinfo{author}{\bibfnamefont{L.}~\bibnamefont{Yu}},
  \bibinfo{author}{\bibfnamefont{C.~A.} \bibnamefont{P\'erez-Delgado}},
  \bibnamefont{and} \bibinfo{author}{\bibfnamefont{J.~F.}
  \bibnamefont{Fitzsimons}}, \bibinfo{journal}{Phys.~Rev.~A}
  \textbf{\bibinfo{volume}{90}}, \bibinfo{pages}{050303}
  (\bibinfo{year}{2014}),
  \urlprefix\url{http://link.aps.org/doi/10.1103/PhysRevA.90.050303}.

\bibitem[{\citenamefont{Newman and Shi}(2017)}]{NS2016}
\bibinfo{author}{\bibfnamefont{M.}~\bibnamefont{Newman}} \bibnamefont{and}
  \bibinfo{author}{\bibfnamefont{Y.}~\bibnamefont{Shi}},
  \bibinfo{journal}{arXiv:1704.07798v1}  (\bibinfo{year}{2017}).

\bibitem[{\citenamefont{Aaronson and Arkhipov}(2011)}]{bib:AaronsonArkhipov10}
\bibinfo{author}{\bibfnamefont{S.}~\bibnamefont{Aaronson}} \bibnamefont{and}
  \bibinfo{author}{\bibfnamefont{A.}~\bibnamefont{Arkhipov}},
  \bibinfo{journal}{Proc. ACM STOC (New York)} p. \bibinfo{pages}{333}
  (\bibinfo{year}{2011}), \eprint{arXiv:1011.3245}.

\bibitem[{\citenamefont{Broome et~al.}(2013)\citenamefont{Broome, Fedrizzi,
  Rahimi-Keshari, Dove, Aaronson, Ralph, and White}}]{bib:broome}
\bibinfo{author}{\bibfnamefont{M.~A.} \bibnamefont{Broome}},
  \bibinfo{author}{\bibfnamefont{A.}~\bibnamefont{Fedrizzi}},
  \bibinfo{author}{\bibfnamefont{S.}~\bibnamefont{Rahimi-Keshari}},
  \bibinfo{author}{\bibfnamefont{J.}~\bibnamefont{Dove}},
  \bibinfo{author}{\bibfnamefont{S.}~\bibnamefont{Aaronson}},
  \bibinfo{author}{\bibfnamefont{T.~C.} \bibnamefont{Ralph}}, \bibnamefont{and}
  \bibinfo{author}{\bibfnamefont{A.~G.} \bibnamefont{White}},
  \bibinfo{journal}{Science} \textbf{\bibinfo{volume}{339}},
  \bibinfo{pages}{794} (\bibinfo{year}{2013}).

\bibitem[{\citenamefont{Spring et~al.}(2013)\citenamefont{Spring, Metcalf,
  Humphreys, Kolthammer, Jin, Barbieri, Datta, Thomas-Peter, Langford, Kundys
  et~al.}}]{bib:spring}
\bibinfo{author}{\bibfnamefont{J.~B.} \bibnamefont{Spring}},
  \bibinfo{author}{\bibfnamefont{B.~J.} \bibnamefont{Metcalf}},
  \bibinfo{author}{\bibfnamefont{P.~C.} \bibnamefont{Humphreys}},
  \bibinfo{author}{\bibfnamefont{W.~S.} \bibnamefont{Kolthammer}},
  \bibinfo{author}{\bibfnamefont{X.-M.} \bibnamefont{Jin}},
  \bibinfo{author}{\bibfnamefont{M.}~\bibnamefont{Barbieri}},
  \bibinfo{author}{\bibfnamefont{A.}~\bibnamefont{Datta}},
  \bibinfo{author}{\bibfnamefont{N.}~\bibnamefont{Thomas-Peter}},
  \bibinfo{author}{\bibfnamefont{N.~K.} \bibnamefont{Langford}},
  \bibinfo{author}{\bibfnamefont{D.}~\bibnamefont{Kundys}},
  \bibnamefont{et~al.}, \bibinfo{journal}{Science}
  \textbf{\bibinfo{volume}{339}}, \bibinfo{pages}{798} (\bibinfo{year}{2013}).

\bibitem[{\citenamefont{Tillmann et~al.}(2013)\citenamefont{Tillmann, Dakic,
  Heilmann, Nolte, Szameit, and Walther}}]{bib:till}
\bibinfo{author}{\bibfnamefont{M.}~\bibnamefont{Tillmann}},
  \bibinfo{author}{\bibfnamefont{B.}~\bibnamefont{Dakic}},
  \bibinfo{author}{\bibfnamefont{R.}~\bibnamefont{Heilmann}},
  \bibinfo{author}{\bibfnamefont{S.}~\bibnamefont{Nolte}},
  \bibinfo{author}{\bibfnamefont{A.}~\bibnamefont{Szameit}}, \bibnamefont{and}
  \bibinfo{author}{\bibfnamefont{P.}~\bibnamefont{Walther}},
  \bibinfo{journal}{Nature Photonics} \textbf{\bibinfo{volume}{7}},
  \bibinfo{pages}{540} (\bibinfo{year}{2013}).

\bibitem[{\citenamefont{Crespi et~al.}(2013)\citenamefont{Crespi, Osellame,
  Ramponi, Brod, Galvao, Spagnolo, Vitelli, Maiorino, Mataloni, and
  Sciarrino}}]{bib:crespi}
\bibinfo{author}{\bibfnamefont{A.}~\bibnamefont{Crespi}},
  \bibinfo{author}{\bibfnamefont{R.}~\bibnamefont{Osellame}},
  \bibinfo{author}{\bibfnamefont{R.}~\bibnamefont{Ramponi}},
  \bibinfo{author}{\bibfnamefont{D.~J.} \bibnamefont{Brod}},
  \bibinfo{author}{\bibfnamefont{E.~F.} \bibnamefont{Galvao}},
  \bibinfo{author}{\bibfnamefont{N.}~\bibnamefont{Spagnolo}},
  \bibinfo{author}{\bibfnamefont{C.}~\bibnamefont{Vitelli}},
  \bibinfo{author}{\bibfnamefont{E.}~\bibnamefont{Maiorino}},
  \bibinfo{author}{\bibfnamefont{P.}~\bibnamefont{Mataloni}}, \bibnamefont{and}
  \bibinfo{author}{\bibfnamefont{F.}~\bibnamefont{Sciarrino}},
  \bibinfo{journal}{Nature Photonics} \textbf{\bibinfo{volume}{7}},
  \bibinfo{pages}{545} (\bibinfo{year}{2013}).

\bibitem[{\citenamefont{Ralph et~al.}(2003)\citenamefont{Ralph, Gilchrist,
  Milburn, Munro, and Glancy}}]{PhysRevA.68.042319}
\bibinfo{author}{\bibfnamefont{T.~C.} \bibnamefont{Ralph}},
  \bibinfo{author}{\bibfnamefont{A.}~\bibnamefont{Gilchrist}},
  \bibinfo{author}{\bibfnamefont{G.~J.} \bibnamefont{Milburn}},
  \bibinfo{author}{\bibfnamefont{W.~J.} \bibnamefont{Munro}}, \bibnamefont{and}
  \bibinfo{author}{\bibfnamefont{S.}~\bibnamefont{Glancy}},
  \bibinfo{journal}{Phys. Rev. A} \textbf{\bibinfo{volume}{68}},
  \bibinfo{pages}{042319} (\bibinfo{year}{2003}),
  \urlprefix\url{http://link.aps.org/doi/10.1103/PhysRevA.68.042319}.

\bibitem[{\citenamefont{Jeong and Kim}(2002)}]{PhysRevA.65.042305}
\bibinfo{author}{\bibfnamefont{H.}~\bibnamefont{Jeong}} \bibnamefont{and}
  \bibinfo{author}{\bibfnamefont{M.~S.} \bibnamefont{Kim}},
  \bibinfo{journal}{Phys. Rev. A} \textbf{\bibinfo{volume}{65}},
  \bibinfo{pages}{042305} (\bibinfo{year}{2002}),
  \urlprefix\url{http://link.aps.org/doi/10.1103/PhysRevA.65.042305}.

\bibitem[{\citenamefont{Tipsmark et~al.}(2011)\citenamefont{Tipsmark, Dong,
  Laghaout, Marek, Je\ifmmode~\check{z}\else \v{z}\fi{}ek, and
  Andersen}}]{PhysRevA.84.050301}
\bibinfo{author}{\bibfnamefont{A.}~\bibnamefont{Tipsmark}},
  \bibinfo{author}{\bibfnamefont{R.}~\bibnamefont{Dong}},
  \bibinfo{author}{\bibfnamefont{A.}~\bibnamefont{Laghaout}},
  \bibinfo{author}{\bibfnamefont{P.}~\bibnamefont{Marek}},
  \bibinfo{author}{\bibfnamefont{M.}~\bibnamefont{Je\ifmmode~\check{z}\else
  \v{z}\fi{}ek}}, \bibnamefont{and}
  \bibinfo{author}{\bibfnamefont{U.}~\bibnamefont{Andersen}},
  \bibinfo{journal}{Phys. Rev. A} \textbf{\bibinfo{volume}{84}},
  \bibinfo{pages}{050301} (\bibinfo{year}{2011}),
  \urlprefix\url{http://link.aps.org/doi/10.1103/PhysRevA.84.050301}.

\bibitem[{\citenamefont{Arrazola and L\"utkenhaus}(2014)}]{PhysRevA.90.042335}
\bibinfo{author}{\bibfnamefont{J.~M.} \bibnamefont{Arrazola}} \bibnamefont{and}
  \bibinfo{author}{\bibfnamefont{N.}~\bibnamefont{L\"utkenhaus}},
  \bibinfo{journal}{Phys. Rev. A} \textbf{\bibinfo{volume}{90}},
  \bibinfo{pages}{042335} (\bibinfo{year}{2014}),
  \urlprefix\url{http://link.aps.org/doi/10.1103/PhysRevA.90.042335}.

\bibitem[{\citenamefont{Braunstein and van Loock}(2005)}]{RevModPhys.77.513}
\bibinfo{author}{\bibfnamefont{S.~L.} \bibnamefont{Braunstein}}
  \bibnamefont{and} \bibinfo{author}{\bibfnamefont{P.}~\bibnamefont{van
  Loock}}, \bibinfo{journal}{Rev. Mod. Phys.} \textbf{\bibinfo{volume}{77}},
  \bibinfo{pages}{513} (\bibinfo{year}{2005}),
  \urlprefix\url{https://link.aps.org/doi/10.1103/RevModPhys.77.513}.

\bibitem[{\citenamefont{Grosshans and Grangier}(2002)}]{PhysRevLett.88.057902}
\bibinfo{author}{\bibfnamefont{F.}~\bibnamefont{Grosshans}} \bibnamefont{and}
  \bibinfo{author}{\bibfnamefont{P.}~\bibnamefont{Grangier}},
  \bibinfo{journal}{Phys. Rev. Lett.} \textbf{\bibinfo{volume}{88}},
  \bibinfo{pages}{057902} (\bibinfo{year}{2002}),
  \urlprefix\url{http://link.aps.org/doi/10.1103/PhysRevLett.88.057902}.

\bibitem[{\citenamefont{Grosshans et~al.}(2003)\citenamefont{Grosshans,
  Van~Assche, Wenger, Brouri, Cerf, and Grangier}}]{GVGWBCG2003}
\bibinfo{author}{\bibfnamefont{F.}~\bibnamefont{Grosshans}},
  \bibinfo{author}{\bibfnamefont{G.}~\bibnamefont{Van~Assche}},
  \bibinfo{author}{\bibfnamefont{J.}~\bibnamefont{Wenger}},
  \bibinfo{author}{\bibfnamefont{R.}~\bibnamefont{Brouri}},
  \bibinfo{author}{\bibfnamefont{N.~J.} \bibnamefont{Cerf}}, \bibnamefont{and}
  \bibinfo{author}{\bibfnamefont{P.}~\bibnamefont{Grangier}},
  \bibinfo{journal}{Nature} \textbf{\bibinfo{volume}{421}},
  \bibinfo{pages}{238} (\bibinfo{year}{2003}),
  \urlprefix\url{http://dx.doi.org/10.1038/nature01289}.

\bibitem[{\citenamefont{Cao et~al.}(2015)\citenamefont{Cao, Zhang, Lo, and
  Ma}}]{bib:Cao2015}
\bibinfo{author}{\bibfnamefont{Z.}~\bibnamefont{Cao}},
  \bibinfo{author}{\bibfnamefont{Z.}~\bibnamefont{Zhang}},
  \bibinfo{author}{\bibfnamefont{H.-K.} \bibnamefont{Lo}}, \bibnamefont{and}
  \bibinfo{author}{\bibfnamefont{X.}~\bibnamefont{Ma}}, \bibinfo{journal}{New
  J.~of Phys.} \textbf{\bibinfo{volume}{17}}, \bibinfo{pages}{053014}
  (\bibinfo{year}{2015}).

\bibitem[{\citenamefont{Barbosa et~al.}(2003)\citenamefont{Barbosa, Corndorf,
  Kumar, and Yuen}}]{PhysRevLett.90.227901}
\bibinfo{author}{\bibfnamefont{G.}~\bibnamefont{Barbosa}},
  \bibinfo{author}{\bibfnamefont{E.}~\bibnamefont{Corndorf}},
  \bibinfo{author}{\bibfnamefont{P.}~\bibnamefont{Kumar}}, \bibnamefont{and}
  \bibinfo{author}{\bibfnamefont{H.}~\bibnamefont{Yuen}},
  \bibinfo{journal}{Phys. Rev. Lett.} \textbf{\bibinfo{volume}{90}},
  \bibinfo{pages}{227901} (\bibinfo{year}{2003}),
  \urlprefix\url{http://link.aps.org/doi/10.1103/PhysRevLett.90.227901}.

\bibitem[{\citenamefont{Nair et~al.}(2006)\citenamefont{Nair, Yuen, Corndorf,
  Eguchi, and Kumar}}]{PhysRevA.74.052309}
\bibinfo{author}{\bibfnamefont{R.}~\bibnamefont{Nair}},
  \bibinfo{author}{\bibfnamefont{H.}~\bibnamefont{Yuen}},
  \bibinfo{author}{\bibfnamefont{E.}~\bibnamefont{Corndorf}},
  \bibinfo{author}{\bibfnamefont{T.}~\bibnamefont{Eguchi}}, \bibnamefont{and}
  \bibinfo{author}{\bibfnamefont{P.}~\bibnamefont{Kumar}},
  \bibinfo{journal}{Phys. Rev. A} \textbf{\bibinfo{volume}{74}},
  \bibinfo{pages}{052309} (\bibinfo{year}{2006}),
  \urlprefix\url{http://link.aps.org/doi/10.1103/PhysRevA.74.052309}.

\bibitem[{\citenamefont{Wilde}(2013)}]{Wilde}
\bibinfo{author}{\bibfnamefont{M.~M.} \bibnamefont{Wilde}},
  \emph{\bibinfo{title}{Quantum Information Theory}}
  (\bibinfo{publisher}{Cambridge University Press}, \bibinfo{year}{2013}),
  \bibinfo{edition}{1st} ed.

\bibitem[{\citenamefont{Ban et~al.}(1997)\citenamefont{Ban, Kurokawa, Momose,
  and Hirota}}]{Ban97}
\bibinfo{author}{\bibfnamefont{M.}~\bibnamefont{Ban}},
  \bibinfo{author}{\bibfnamefont{K.}~\bibnamefont{Kurokawa}},
  \bibinfo{author}{\bibfnamefont{R.}~\bibnamefont{Momose}}, \bibnamefont{and}
  \bibinfo{author}{\bibfnamefont{O.}~\bibnamefont{Hirota}},
  \bibinfo{journal}{International Journal of Theoretical Physics}
  \textbf{\bibinfo{volume}{36}}, \bibinfo{pages}{1269} (\bibinfo{year}{1997}).

\bibitem[{\citenamefont{Aharonov et~al.}(1993)\citenamefont{Aharonov,
  Davidovich, and Zagury}}]{bib:ADZ}
\bibinfo{author}{\bibfnamefont{Y.}~\bibnamefont{Aharonov}},
  \bibinfo{author}{\bibfnamefont{L.}~\bibnamefont{Davidovich}},
  \bibnamefont{and} \bibinfo{author}{\bibfnamefont{N.}~\bibnamefont{Zagury}},
  \bibinfo{journal}{Phys. Rev. A} \textbf{\bibinfo{volume}{48}},
  \bibinfo{pages}{1687} (\bibinfo{year}{1993}).

\bibitem[{\citenamefont{Rohde et~al.}(2011)\citenamefont{Rohde, Schreiber, {\v
  S}tefa{\v n}{\' a}k, Jex, and Silberhorn}}]{bib:RohdeSchreiber10}
\bibinfo{author}{\bibfnamefont{P.~P.} \bibnamefont{Rohde}},
  \bibinfo{author}{\bibfnamefont{A.}~\bibnamefont{Schreiber}},
  \bibinfo{author}{\bibfnamefont{M.}~\bibnamefont{{\v S}tefa{\v n}{\' a}k}},
  \bibinfo{author}{\bibfnamefont{I.}~\bibnamefont{Jex}}, \bibnamefont{and}
  \bibinfo{author}{\bibfnamefont{C.}~\bibnamefont{Silberhorn}},
  \bibinfo{journal}{New J. Phys.} \textbf{\bibinfo{volume}{13}},
  \bibinfo{pages}{013001} (\bibinfo{year}{2011}).

\bibitem[{\citenamefont{Gerry and Knight}(2004)}]{GerryKnight}
\bibinfo{author}{\bibfnamefont{C.}~\bibnamefont{Gerry}} \bibnamefont{and}
  \bibinfo{author}{\bibfnamefont{P.}~\bibnamefont{Knight}},
  \emph{\bibinfo{title}{Introductory Quantum Optics}}
  (\bibinfo{publisher}{Cambridge University Press}, \bibinfo{year}{2004}),
  \bibinfo{edition}{1st} ed.

\bibitem[{\citenamefont{Bartlett et~al.}(2002)\citenamefont{Bartlett, Sanders,
  Braunstein, and Nemoto}}]{PhysRevLett.88.097904}
\bibinfo{author}{\bibfnamefont{S.~D.} \bibnamefont{Bartlett}},
  \bibinfo{author}{\bibfnamefont{B.~C.} \bibnamefont{Sanders}},
  \bibinfo{author}{\bibfnamefont{S.~L.} \bibnamefont{Braunstein}},
  \bibnamefont{and} \bibinfo{author}{\bibfnamefont{K.}~\bibnamefont{Nemoto}},
  \bibinfo{journal}{Phys. Rev. Lett.} \textbf{\bibinfo{volume}{88}},
  \bibinfo{pages}{097904} (\bibinfo{year}{2002}),
  \urlprefix\url{http://link.aps.org/doi/10.1103/PhysRevLett.88.097904}.

\bibitem[{\citenamefont{Yurke and Stoler}(1986)}]{YS1986}
\bibinfo{author}{\bibfnamefont{B.}~\bibnamefont{Yurke}} \bibnamefont{and}
  \bibinfo{author}{\bibfnamefont{D.}~\bibnamefont{Stoler}},
  \bibinfo{journal}{Phys. Rev. Lett.} \textbf{\bibinfo{volume}{57}},
  \bibinfo{pages}{13} (\bibinfo{year}{1986}),
  \urlprefix\url{http://link.aps.org/doi/10.1103/PhysRevLett.57.13}.

\bibitem[{\citenamefont{Rohde et~al.}(2015)\citenamefont{Rohde, Motes, Knott,
  Fitzsimons, Munro, and Dowling}}]{RMKFMD15}
\bibinfo{author}{\bibfnamefont{P.~P.} \bibnamefont{Rohde}},
  \bibinfo{author}{\bibfnamefont{K.~R.} \bibnamefont{Motes}},
  \bibinfo{author}{\bibfnamefont{P.~A.} \bibnamefont{Knott}},
  \bibinfo{author}{\bibfnamefont{J.}~\bibnamefont{Fitzsimons}},
  \bibinfo{author}{\bibfnamefont{W.~J.} \bibnamefont{Munro}}, \bibnamefont{and}
  \bibinfo{author}{\bibfnamefont{J.~P.} \bibnamefont{Dowling}},
  \bibinfo{journal}{Phys. Rev. A} \textbf{\bibinfo{volume}{91}},
  \bibinfo{pages}{012342} (\bibinfo{year}{2015}),
  \urlprefix\url{http://link.aps.org/doi/10.1103/PhysRevA.91.012342}.

\end{thebibliography}

\appendix

\section{Calculation of eigenvalues}\label{app:eigen}

Let $|x\>$ and $|y\>$ be normalized states that are not orthogonal to one another. Let $|z\>$ be a normalized state that is orthogonal to $\ket{x}$ and in the plane spanned by $\ket{x}$ and $\ket{y}$. One can write $|y\>$ in terms of $\ket{x}$ and $\ket{z}$ as
\begin{align}
|y\> 
&= ( |x\>\<x| + |z\>\<z|) |y\> \notag\\
 &= |x\>\cos \theta + |z\>\sin \theta \ ,
\end{align}
where $\cos \theta=\<x|y\>$ and $\sin\theta=\<z| y\>$. Then a given matrix
\begin{align}
M= |x\>\<x| - C |x\>\<y| - C |y\>\<x| + |y\>\<y|.
\end{align}
can be rewritten in terms of $\ket{x}$ and $\ket{z}$ as
\begin{align}
M =&
|x\>\<x| (1- 2C \cos \theta + \cos^2 \theta ) \nonumber\\
& + |x\>\<z| (-C \sin \theta + \sin \theta \cos \theta) \notag\\
&
+|z\>\<x| (-C \sin \theta + \sin \theta \cos \theta) \nonumber\\
&+ |z\>\<z| \sin^2 \theta ,
\end{align}
for which its eigenvalues are $\lambda_\pm=(1\pm C)(1\mp \cos \theta)$.

When $M=\hat{O}_k$, $C=A_k$ and $\cos\theta=\braket{\tilde{g}_k|\tilde{h}_k}=A_k$, we have $\lambda_+=\lambda_-=1-A_k^2$. \
When $M=\hat{Q}$, $C=B$ and $\cos\theta=\braket{\psi_{\bf x}|\psi_{\bf 0}}=B$, we have $\lambda_+=\lambda_-=1-B^2$.

\section{Calculation of $q_k$ and $A_k$ in the limit $d\rightarrow\infty$}\label{app:deriv}

In the limit $d\rightarrow \infty$, we can drop the modulus in $\phi({\bf z})$ and use the multinomial theorem to simplify $q_k$ and $A_k$. We have
\begin{align}
q_k\stackrel{d\rightarrow\infty}{=} &\sum_{\substack{{\bf z}\in \mathbb{N}^m\\ z_1+\ldots+ z_m=k}}e^{-m|\alpha|^2}\frac{|\alpha|^{2(z_1+\ldots +z_m)}}{z_1!z_2!\ldots z_m!}\nonumber\\
=& \sum_{\substack{{\bf z}\in \mathbb{N}^m\\ z_1+\ldots +z_m=k}}e^{-m|\alpha|^2}\frac{|\alpha|^{2k}}{k!} {{k}\choose{z_1!z_2!\ldots z_m!}}\nonumber\\
=& e^{-m|\alpha|^2}\frac{(m|\alpha|^2)^k}{k!} \ ,
\end{align}
where ${{k}\choose{z_1,z_2,\ldots ,z_m}}:=\frac{k!}{z_1! z_2! \ldots z_m!}$ is the multinomial coefficient and
\begin{align}
A_k\stackrel{d\rightarrow\infty}{=}& \frac{1}{q_k}\sum_{\substack{{\bf z}\in \mathbb{N}^m\\ z_1+\ldots +z_m=k}}e^{-m|\alpha|^2}\frac{|\alpha|^{2(z_1+\ldots+z_m)}}{z_1!\ldots z_m!}(-1)^{{\bf x}\cdot {\bf z}}\nonumber\\
=&  \frac{1}{q_k}\sum_{\substack{{\bf z}\in \mathbb{N}^m\\ z_1+\ldots+ z_m=k}}e^{-m|\alpha|^2}\frac{|\alpha|^{2k}}{k!}{{k}\choose{z_1,z_2,\ldots ,z_m}}(-1)^{{\bf x}\cdot {\bf z}}\nonumber\\
=& \frac{1}{q_k}e^{-m|\alpha|^2}\frac{|\alpha|^{2k}}{k!}\left((-1)^{x_1}+\ldots+ (-1)^{x_m}\right )^k\nonumber\\
=&\frac{1}{q_k}(m-2{\rm wt}({\bf x}))^k e^{-m |\alpha|^2}\frac{|\alpha|^{2k}}{k!} \ ,
\end{align}
respectively, where ${\rm wt}({\bf x})$ is the Hamming weight of ${\bf x}$.

\end{document}